\documentclass[aps,prb,superscriptaddress,twocolumn]{revtex4}
\usepackage{graphicx}
\usepackage{dcolumn}
\usepackage{color}
\usepackage{soul}
\usepackage[breaklinks,colorlinks = true,linkcolor = blue,urlcolor=blue,citecolor=blue]{hyperref}
\usepackage{amsmath}

\begin{document}
\title{Giant caloric effects in spin chain materials}
\author{A.A. Zvyagin}
\affiliation{B. Verkin Institute for Low Temperature Physics and Engineering of the National Academy of Sciences of Ukraine, Nauky Ave., 47, Kharkiv, 61103, Ukraine}
\affiliation{Max-Planck-Institut f\"ur Physik komplexer Systeme, N\"othnitzer
Str., 38, D-01187, Dresden, Germany}
\author{V.V. Slavin}
\affiliation{B. Verkin Institute for Low Temperature Physics and Engineering of the National Academy of Sciences of Ukraine, Nauky Ave., 47, Kharkiv, 61103, Ukraine}

\begin{abstract}
The giant electro- and elasto-caloric effects in spin chain materials are predicted. The theory is based on the exact quantum mechanical solution of the problem. It is shown that the giant jumps in the entropy and the temperature caused by the caloric effect are weakly affected by the initial temperature. The effect can be used for the cooling of new quantum devices (like systems of qubits in quantum computers). On the other hand, since large changes are predicted in the narrow neighborhood of the critical point, the predicted effect can be used in ultra-sensitive electric and stress sensors for modern microelectronics.    
\end{abstract}

\date{\today}
\maketitle

\section{Introduction}
According to the data of the International Institute of Refrigeration, nowadays cooling demands about 17\% of the total used energy \cite{CDP,I,E}. The prompt urbanization implies that the part of the energy consumed for the refrigeration will grow considerably \cite{I}. Therefore, the quest for novel refrigeration techniques is among the most important goals of the modern energy science. The standard approach for cooling is the vapour compression. In that method the refrigeration is the result of expansion of previously compressed gaseous refrigerant, repeated in a cyclic way. That way of cooling is widely used, being well developed, and relatively high effective. On the other hand, the unavoidable losses of the energy caused by, e.g., the super-heating of the compressing refrigerant, or the expansion in the valve, controlling the flow of the gas stream, etc. demand the search of alternative cooling technologies. Among various approaches, caloric refrigeration is the promising one \cite{E}, especially for a low temperature usage. The nature of magneto-, electro- and mechano-caloric effects 
\cite{Moya_Mathu,Reis_Ma} is in the reduction of the temperature of a system subjected to an adiabatic external magnetic or electric field, and for the mechano-caloric effect an applied pressure (baro-calorics), or a uni-axial strain (elasto-calorics). Similar to the vapour compression method, the caloric effect can be used in the thermodynamic cycle \cite{KPTP}. While the application of the caloric effect for cooling of large industrial subjects is still not commercially viable, it can be used for the refrigeration of supercomputers based on superconducting devices, or quantum computers, which need low temperatures. From the pure research side, the caloric effect can be applied for reaching ultra-low temperatures, which cannot be obtained using other methods.    

Magneto-, electro- and mechano-caloric effects are the manifestation of laws of thermodynamics. For systems in thermodynamic equilibrium adiabatic variations of governing external parameters (magnetic and electric fields or stress) produce changes of the temperature and entropy of the system. The largest changes of the temperature and entropy are expected at phase transitions, where properties of matter are renormalized drastically under the action of relatively small variations of external parameters. In low-dimensional systems fluctuations at phase transitions are strengthened comparing to the standard tree-dimensional ones. However, the precise role of low dimensional enhancement in caloric effects is not well understood yet. 

In this study we show that electro- and elasto-caloric effects in quantum spin chain materials can be colossal. We show that despite relatively weak couplings between spin, electric and elastic subsystems, the changes of the temperature and entropy in mentioned caloric effects can be very large, of order of the strongest value of the exchange integral for the spin-spin coupling along the distinguished direction. Our results demonstrate how the electric field- and strain-induced spin nematicity in the vicinity of the quantum phase transition affects the entropy. We show that huge jumps of the entropy and the temperature caused by the caloric effects in quantum spin chain materials are weakly affected by the temperature, due to strong quantum fluctuations. The predicted giant caloric effects can be harnessed for cooling down to ultracold temperatures. We anticipate that those effects can be adopted for cooling of modern supercomputers (e.g., superconducting ones), and of ensembles of quantum bits (qubits) in quantum computers, where low temperatures provide the reduction of noise and quantum decoherence. Also the predicted effects can be used for production of ultra-sensitive sensors for the electric field and mechanical stress.

\section{Caloric effect}
From thermodynamics we know that the isothermal change of the entropy ${\cal S}$ of the system in equilibrium under adiabatic changes of external parameter $a$ at the fixed internal thermodynamic variable $b$, thermodynamically conjugated to $a$, can be written as   
\begin{equation}
\Delta {\cal S} (a_1 \to a_2) = -\int_{a_1}^{a_2} d a \left(\frac {\partial b} {\partial T}\right)_a \ , 
\end{equation}
where $T$ is the temperature. For example, for the magneto-caloric effect $a$ is the external magnetic field $H$ and $b$ is the magnetization of the system $M$, for the electro-caloric effect $a$ is the external electric field $E$ and $b$ is the electric induction $D$, and for the elasto-caloric effect the stress $\sigma$ plays the role of $b$, and the strain $\epsilon$ is $a$. The Maxwell relation implies that in the thermal equilibrium $\alpha_a = (\partial b)/(\partial T)|_a = (\partial {\cal S})/(\partial a)|_T$. The value $\alpha_a$ is the expansion coefficient, related to the variation of $a$. The change of the temperature for the caloric effect, in turn, can be presented as 
\begin{equation}
\Delta {\cal T} (a_1 \to a_2) = \int_{a_1}^{a_2} d a \frac {T}{(a_2-a_1)c_a}\left(\frac {\partial b }{\partial T}\right)_a \ , 
\label{Delta_T}
\end{equation}
where $c_a$ is the specific heat at the fixed value of $a$. For the quantitative description of the caloric effect one uses the Gr\"uneisen ratio 
\begin{equation}
\Gamma_a = \frac{\left(\frac{\partial {\cal S}}{\partial a}\right)_{T}}{c_a}  = \frac{\alpha_a}{c_a}  \,  
\label{Gamma_a}
\end{equation}
which was originally introduced long time ago \cite{Gr} for the Einstein model to study the quantitative characteristics of the effect of volume change of a crystal lattice on its vibrational frequencies. The Gr\"uneisen ratio determines the renormalization of the entropy under adiabatic changes of $a$ for the caloric effect. 

It is clear from those equations that to get the maximal caloric refrigeration effect one needs to use systems with the small specific heat, and systems, in which the changes of the internal thermodynamic variable $b$ are the largest for the variation of the external parameter $a$ from $a_1$ to $a_2$. The natural conditions for the latter is to use systems near phase transitions, governed by the external parameter $a$. It is well known that some phase transitions are characterized by the drastic changes of the state of the system under relatively small variations of the external parameters. For the small specific heat it is natural to chose the quantum system with excitations with activation. For those systems the specific heat is exponentially small at temperatures lower than the value of the energy gap, necessary to activate an excited state. 

\section{Caloric Effects in spin chains}
Among other systems, quantum one-dimensional spin 1/2 ones play the special role. (Quasi) one-dimensional systems are ones, in which an interaction between ions in one space direction is much stronger than in the other directions. First, for those systems fluctuations (which provide phase transformations) are enhanced comparing to the usual three-dimensional ones due to the peculiarities in the density of states. Those fluctuations often destroy ordering in one-dimensional systems \cite{MW,H}. Then, for quantum one-dimensional systems one can use the great number of theoretical results, often exact. Such a results are mostly unavailable for three-dimensional systems. Last but not least, recent technological progress in fabrication of novel materials with demanded properties permitted to obtain numerous systems with one-dimensional properties, including spin chain materials.   

Consider the spin-1/2 chain system with the Hamiltonian 
\begin{equation}
{\cal H}_0 = \sum_n[J(S_n^xS_{n+1}^x + S_n^yS_{n+1}^y) + J^z S_n^zS_{n+1}^z]  .
\label{H1}
\end{equation}
Here $S_n^{x,y,z}$ are the operators of the components of spins 1/2 situated at the site $n$ of the chain, $J$ are the exchange constants, $J_z \ne J$ determines the magnetic anisotropy of the exchange interaction. Thermodynamic characteristics of the system with the Hamiltonian ${\cal H}_0$ do not depend on the sign of $J$ \cite{LSM}. The free energy of the system is $F(|J_z|,T)$ for $J_z >0$ i.e. for the antiferromagnetic interactions, and it is $-F(|J_z|,-T)$ for $J_z <0$ i.e. for the ferromagnetic interactions. Then for convenience we introduce the following notations: $\Delta = J_z/J$ and $t = k_BT/J$ (we do not study the time dependence in our work, hence there must be no confusion with the standard notations for time), where $k_B$ is the Boltzmann constant. It is known that in the ground state ($T=0$) quantum phase transitions take place in the system. For $\Delta < -1$ the chain is ferromagnetically ordered; for $\Delta >1$ it is ordered antiferromagnetically, and for $-1 < \Delta < 1$ the system is disordered \cite{Zvb,Kl2}. It means that $\Delta =\pm 1$ are quantum critical points \cite{John}. All eigenvalues of the considered quantum system can be obtained exactly using the Bethe ansatz \cite{Zvb}. Performing exact calculations for the free energy of the system (see Appendix \ref{AppendixA}) we obtain necessary results for the entropy of the spin chain for calculations of the caloric effects.  

Magneto-caloric effect for such a system is well studied theoretically, see, e.g., \cite{magcal,magcal1}. Here we concentrate on the elasto-caloric and electro-caloric effects, where the former is associated with a strain and the latter on including an external electric field in a linear
approximation. First, let us consider the strain $\epsilon_{zz}$. It produces shifts of the magnetic ions themselves, or neighboring nonmagnetic ions (ligands) involved into the indirect exchange coupling \cite{Lut}. In the main linear approximation such a strain renormalizes the exchange parameter of the spin chain as $J_z \to J_z(1- f_{zz}\epsilon_{zz})$, where $f_{zz} = (\partial J_z/\partial R_z)$ is the component of the tensor of the magneto-elastic interaction, $R_z$ is the component of the co-ordinate of the magnetic ion. The electro-caloric effect can take place in crystals, in which magnetic ions are situated not in the center of inverse \cite{Bl}. The external electric field in such a system shifts ions one with respect to other. It produces the change of the effective exchange interaction between magnetic ions. In the linear approximation the electric field $E_i$ ($i=x,y,z$) renormalizes the exchange parameter of the spin chain as $J_z \to J_z(1- h_{i zz}E_{i})$, where $h_{izz}$ is the component of the tensor of the electro-magnetic interaction. The external electric field or the distortion change the population imbalance for spins and results in the cooling of the system. 
    
\begin{figure}[ht]
\center{\includegraphics[width=6cm]{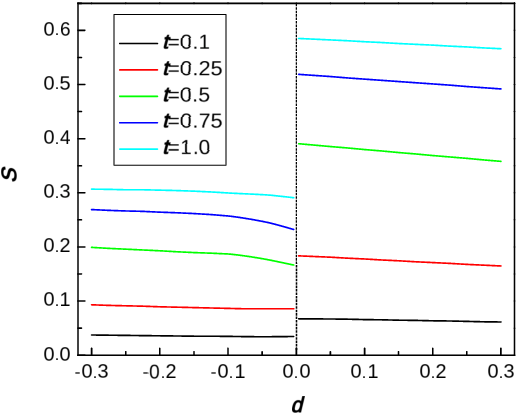}}
\caption{The dependence of the entropy ${\cal S}$ of the spin chain per site on the magnetic anisotropy parameter $d = \Delta - {\rm sign} (\Delta )$ at different values of dimensionless temperature $t=k_BT/J$.}
\label{fig1}
\end{figure}

\begin{figure}[ht]
\center{\includegraphics[width=6cm]{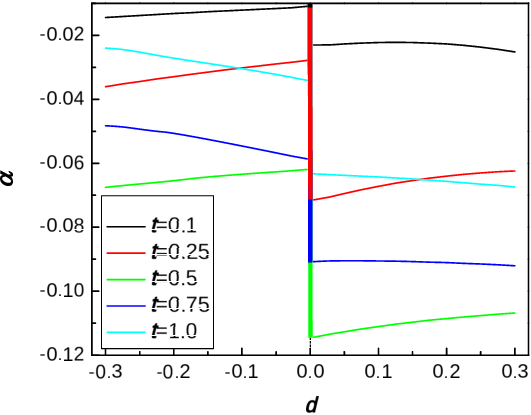}} 
\caption{The dependence of the expansion parameter $\alpha_d$ of the spin chain per site on the magnetic anisotropy parameter $d$ at different values of temperature $t$.
Wide vertical lines demonstrate schematically $\delta$-function like feature at $d=0$.}
\label{fig2}
\end{figure}

Considered magneto-elastic and electro-magnetic couplings renormalize only one parameter of the spin Hamiltonian, $J_z =J\Delta $.
For convenience we introduce the parameter $d = \Delta - {\rm sign} (\Delta )$, which determines the deviation from the isotropic exchange interaction, i.e. from quantum critical points caused by the external electric field $E_i$ or the strain $\epsilon_{zz}$. From now on we consider the antiferromagnetic chain $\Delta >0$ (the ferromagnetic case can be studied analogously). The parameter $d$ is an effective field which acts on the component of the inter-ion spin quadrupole order parameter, responsible for the spin nematicity of the system \cite{ZS}. To remind, unlike a spin dipole order parameter, which is the vector, the spin nematic order parameter is a director (similar to the situation in organic liquid crystals \cite{dG}), and breaks the rotational symmetry in the system. In what follows we will consider $d$ as the external governing parameter $a$, since it depends on the strain $\epsilon_{zz}$ or the electric field $E_i$. Obviously one has $(\partial/\partial E_i) = -h_{i zz}(\partial /\partial d)$ and 
$(\partial/\partial \epsilon_{zz}) = -f_{zz}(\partial /\partial d)$. The component of the inter-site spin quadrupole moment $\langle S_n^zS_{n+1}^z \rangle - (1/3)\langle {\bf S}_n\cdot {\bf S}_{n+1} \rangle $ plays the role of the internal parameter $b$, conjugated to $d$ \cite{ZS} (the role of $a$ plays the anisotropy  parameter $d$,  i.e. ${\cal S}_{a}\to {\cal S}_{d}$, $c_a\to c_d, \alpha_a \to \alpha_d$ and $\Gamma_a\to \Gamma_d$). For $d=0$ the spin nematicity is zero.  

\section{Results}
\begin{figure}[ht]
\center{\includegraphics[width=6cm]{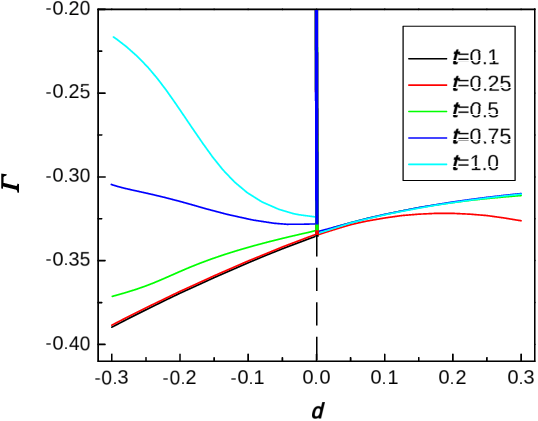}}
\caption{The dependence of the Gr\"uneisen parameter $\Gamma_a$ (see (\ref{Gamma_a}) with $a=d$) of the spin chain per site on the magnetic anisotropy parameter $d$ at different values of temperature $t$.
Wide vertical lines demonstrate schematically $\delta$-function like feature at $d=0$.}
\label{fig3}
\end{figure}

\begin{figure}[ht]
\center{\includegraphics[width=6cm]{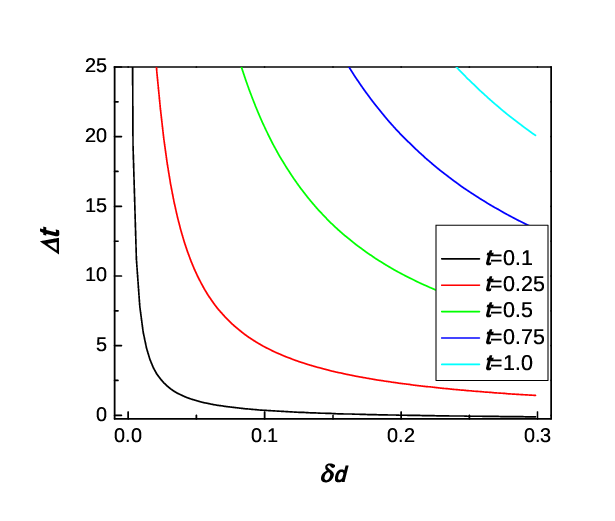} \\a)}
\center{\includegraphics[width=6cm]{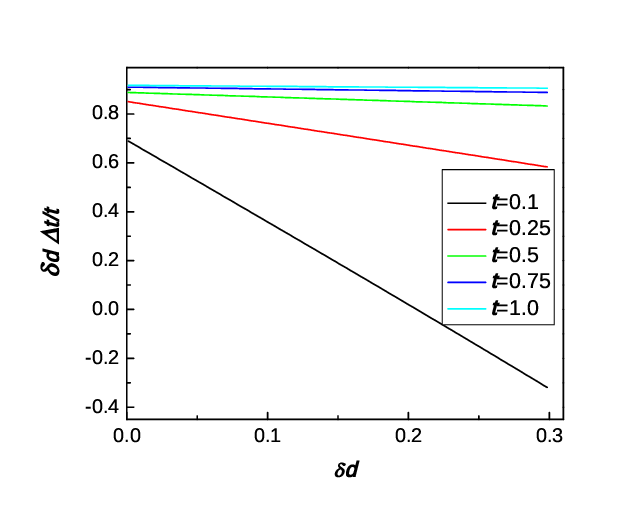} \\b)}
\caption{a) The dependencies of the temperature change $\Delta t$ (see (\ref{Delta_T}) with $a=d$) per site of the spin chain on the change of magnetic anisotropy parameter $\delta d$ at different values of temperature $t$. In the considered region of $d\in(0,0.3)$ the value of $\Delta t$ changes its sign in the case of the initial temperature $t=0.1$ only.
b) The same dependencies in variables $(\delta d \frac{\delta t}{t}, \delta d)$. It can be seen that the magnitude of the caloric effect is from $\approx 0.7$ for $t=0.1$  up to $\approx 0.85$  for $t=1.0$. }
\label{fig4}
\end{figure}

The main results are presented in Fig.~\ref{fig1} ---  Fig.~\ref{fig4}. First, the calculation of the entropy of the spin chain per site as a function of $d$ for various values of temperature from the nonlinear integral equations, which exactly describe thermodynamics (see Appendix \ref{AppendixA}) manifests jumps in ${\cal S}(d)$ at $d=0$ for $T \ne 0$, see Fig.~\ref{fig1}. This is why, the expansion coefficient $\alpha_d$ per site, which is the derivative of ${\cal S}(d)$ with respect to $d$, and the Gr\"uneisen ratio $\Gamma_d$ per site, proportional to $\alpha_d$, manifest the $\delta$-function like feature at $d=0$, see Fig.~\ref{fig2}, and Fig.~\ref{fig3}, respectively. Fig.~\ref{fig4} presents the results for the change of the temperature per site $\Delta t$,  caused by the adiabatic change of $d$. All these dependencies were obtained for several values of the temperature: $t=T/J=0.1, 0.25, 0.5, 0.75, 1$. We stress that the results are exact. 

The jump of the entropy (which grows with the temperature) causes the features of the caloric effect in the quantum spin chain material: The changes of the expansion parameter and the Gr\"uneisen ratio (Fig.~\ref{fig2} and Fig.~\ref{fig3}) at $d=0$ are huge (of order of the exchange constant). They exist for a large temperature range, not only in the ground state, as for magneto-caloric effect for spin chains \cite{magcal,magcal1}. Even the large value of the specific heat cannot overcome the growth of the Gr\"uneisen ratio. The increase of the temperature (to remind, the temperature in our calculations is measured in the units of the exchange integral) does not remove the giant renormalization of the entropy at the quantum critical point, and does not lead to broadening of the entropy jump. It is the manifestation of quantum fluctuations enhancement by the quantum critical point at nonzero temperatures \cite{Sach}. The temperature renormalization for the caloric effect (see Fig.~\ref{fig4}) is of the order of the exchange constant (the values for the temperature renormalization $\Delta t$ obviously do not depend on the parameters of the magneto-elastic and electro-magnetic couplings). Notice that according to \cite{class} and references therein, the term "giant" is attached to the materials with the first order phase transition. In our case one has the jump of the entropy, the first derivative of the thermodynamic potential, hence we use the term "giant" for the studied effect. We can see from Fig.~\ref{fig4} that the renormalization of the temperature caused by the considered caloric effect can be of order of the initial temperature.

Now we present some analytic results for the caloric effect in the quantum antiferromagnetic spin chain material. Using the known series (see Appendix \ref{AppendixB}) at low temperatures $T \ll J, J_z$ for $d < 0$ 
($\Delta <1$) we get for the entropy per site ${\cal S}|_{d \to -0} =t/3 + \dots$. The specific heat per site is $c_d|_{d \to -0} = t/3$. The expansion coefficient per site is $\alpha_d =\partial {\cal S}/\partial d$; we obtain $\alpha_d|_{d \to -0} = t/9$. Then the Gr\"uneisen ratio is $\Gamma_d|_{d \to -0} = -1/3$. For both negative $d_{1,2}$ the change of the temperature caused by the caloric effect is 
\begin{equation}
\Delta t  = \frac{t}{|\delta d|}\int_{d_1}^{d_2} \Gamma_d(x) dx \approx \frac{t}{3|\delta d|}
\end{equation} 
($\delta d =(d_2-d_1)$; here and below we integrate with respect to $d$, and replace the variable $d$ by $x$ for convenient reading).  For $d > 0$ at low temperatures for $d \to +0$ the gap is small and the main contribution comes from the pre-exponential multipliers. For the specific heat, expansion coefficient and the Gr\"uneisen ratio we obtain $c_d|_{d \to +0} = 2t/3$, $\alpha_d|_{d \to +0} = 2t/9$, and $\Gamma_d|_{d \to +0} = -1/3$. For both positive $d_{1,2}$ the change of the temperature caused by the caloric effect is 
\begin{equation}
\Delta t  = \frac{t}{\delta d}\int_{d_1}^{d_2} \Gamma_d(x) dx \approx \frac{t}{3(\delta d)} \ .
\end{equation}  

The most interesting case corresponds to the situation in which $d_1 < 0$ and $d_2 > 0$ or vice versa, i.e., the transition across the quantum critical point. 
According to Eq.~(\ref{FreeEnergy1}) and Eq.~(\ref{gapless}) of the Appendix 
at low temperature limit and in small neighborhood of $d=0$ the specific heat is $c_d = t \frac{\partial {\cal S}}{\partial t} \approx {\cal S}$, and the Gr\"uneisen ratio is  $\Gamma_d(x) = \frac{1}{c_d}\frac{\partial{\cal S}}{\partial d}  \approx \frac{1}{\cal S} \frac{\partial{\cal S}}{\partial d} = \frac{\partial\ln {\cal S}}{\partial d}$. It means that the integral of $\Gamma_d(x)$ over this small neighborhood of $d=0$ (i.e. $0\in (-\delta d/2,\delta d/2)$, 
$|\delta d|\ll 1$) is $\int_{-\delta d/2}^{\delta d/2}\Gamma_d(x)dx = \ln 2$, and the change of the temperature caused by the caloric effect is
\begin{equation}
\Delta t  = \frac {t}{\delta d} \int_{-\delta d/2}^{\delta d/2} \Gamma_d(x) d x \approx \frac{t \ln 2}{\delta d} \ . 
\label{dT}
\end{equation}

Unlike most of quantum phenomena, the discovered caloric effect manifests itself not only in the low-temperature region. Moreover, as can be seen from Fig.~\ref{fig1}, with increasing temperature, the
jump in entropy increases, and, consequently, the magnitude of the effect itself. On the one hand, it simplifies the experimental detection of the phenomenon, on the other hand, it removes the question of the adequacy of the applicability of the one-dimensional spin model. Indeed, strictly one-dimensional magnetic chains are absent in nature due to the presence of arbitrarily weak but finite inter-chain interactions. Wherein, the effects associated with such interactions (i.e. the manifestation of quasi-one-dimensionality effects) appear only at temperatures below the characteristic energies of inter-chain interactions. In our case, this means that it is always possible to choose a temperature range at which the discovered caloric effect is fully manifested, and the effects associated with quasi-one-dimensionality are negligible small. The predicted caloric effect can be also manifested in organic spin chain materials: There vibrations of molecules produce the anisotropy of spin-spin interactions, similar to strains in crystals.  

Let us estimate the value of the effect. There exists a large variety of spin chain materials with the exchange constants from $J \sim 2100$~K for SrCuO$_2$ (with the magnetic ordering temperature below 2~K) \cite{SrCu}, CaCu$_2$O$_3$ with $J \sim 2000$~K and ordering temperature $\sim 25$~K \cite{CaCu}, InCuPO$_5$ with $J \sim 100$~K and the inter-chain coupling of order of 0.07$J$ \cite{InCu}, Cs$_4$CuSb$_2$Cl$_{12}$ with $J \sim 180$~K, which manifests the Heisenberg spin chain behavior down to 0.7~K \cite{CsCu}, KCuF$_3$ with $J \sim 400$~K \cite{KCu} (with the ordering temperature $\sim 40$~K) to organic spin chain complexes like Cu(C$_4$H$_4$N$_2$)(NO$_3$)$_2$ with the intra-chain exchange $J \sim 10$~K and very small inter-chain interactions \cite{cupyr}. For example, the low-temperature change of the temperature caused by the predicted caloric effect according to Eq.~(\ref{dT}) can be equal to the initial temperature for the changes of the anisotropy parameter $\delta d \sim 0.3$, which is of order of the known values of the magnetic anisotropy caused by the spin-elastic coupling in spin chain materials \cite{sl1,sl2,sl3,ZS}. The value $|\delta d|\Delta t$ almost reaches the value $\ln 2$ at low temperatures, i.e. the entropy of the spin 1/2, and it weakly grows with the growth of the initial temperature. To compare, the temperature change for the magneto-caloric effect in organic spin chain system [Cu($\mu$-C$_2$O$_4$)(4-aminopyridine)2(H$_2$O)]$_n$ ($J \sim 3.2$~K) was of order of 1.2~K for the change of the magnetic field from 4 to 7~T \cite{Cupy}, and in the spin chain material Cu(NO$_3$)$_2$ 2.5 H$_2$O ($J \sim 5$~K) it was about 1~K for the change of the external magnetic field from zero to 3~T \cite{CuN}. As for the comparison with the electro-caloric effect in spin chain materials with the second order quantum phase transition, the recent study \cite{Z} calculated the value of the renormalization of the Gr\"uneisen ratio at the critical point of order of 1.6, and it strongly decays with the growth of the temperature, cf. Fig.~\ref{fig3} where the renormalization of the Gr\"unesen parameter is much larger, and it is almost temperature independent near $d=0$. 

\section{Summary}
In summary, we have predicted the giant electro- and elasto-caloric effect in the spin chain material. The predicted caloric effect can be used for cooling of modern supercomputers and small quantum systems, e.g., one-dimensional arrays of qubits (quantum registers) in quantum computers. Since the giant effect manifests itself even in a small neighborhood of $d=0$ point, this makes it possible to use this phenomena to create ultra-sensitive sensors for electric field and mechanical stress (both compression and rarefaction). To do this one needs the antiferromagnetic spin-1/2 chain material with the isotropic (or almost isotropic) exchange coupling along the spin chain. Then adiabatically applying the external electric field or external uniaxial stress one either shifts the system to the spin nematic anisotropic phase or vice versa.

\section*{ACKNOWLEDGEMENTS}
A.A.Z. acknowledges the support from the Deutsche Forschungsgemeinschaft under SFB 1143 (Project-ID No.
247310070) and from N.I. Akhiezer Foundation. V.V.S. acknowledges the support from the Project IMPRESS-U: N2401227.

\appendix
\section{Exact results for the thermodynamics of the spin-1/2 chain}
\label{AppendixA}

Consider the Hamiltonian of the chain system with the nearest neighbour interactions between spins 1/2 
\begin{eqnarray}
&&{\cal H}_0 = \sum_n[J(S_n^xS_{n+1}^x + S_n^yS_{n+1}^y) + J^z S_n^zS_{n+1}^z] - \nonumber \\ 
&&g\mu_BH\sum_nS_n^z .
\label{A1}
\end{eqnarray}
Here $H$ is the external magnetic field,  $g$ is the effective $g$-factor, and $\mu_B$ is Bohr's magneton. It is convenient to introduce the following notation $h=g\mu_BH/J$. 

We are interested in the properties of the system at $T \ne 0$. The most convenient method to describe thermodynamics of the spin chain is quantum transfer matrix one \cite{Kl1}. To be concrete, let us study in detail the case $-1 \le \Delta \le 1$, i.e. the easy-plane magnetic anisotropy. It is convenient to define $\Delta = \cos \theta$. Consider $R_{a_i b_i}^{m_i m_{i+1}}(u)$, the standard $R$ matrix of the spin-1/2 chain with the easy-plane magnetic anisotropy \cite{Zvb}. The nonzero matrix elements of that $R$ matrix are $R_{12}^{21} = R_{21}^{12} =1$, $R_{11}^{11} = R_{22}^{22} = \sin [(\theta(u+2)/2]/ \sin (\theta)$, and $R_{12}^{12} = R_{21}^{21} = \sin [(\theta u)/2]/ \sin (\theta)$, where the index 1 is related to the state with spin up, and the index 2 is related to the state with spin down. Here $u$ is the spectral parameter. The indices $a_i$ and $b_i$ denote states in the Hilbert space of the spin at site $i$, and $m$ denotes states in the auxiliary Hilbert space. Let us construct the row-to-row transfer matrices $\tau^b_a(u)$ as the trace over the auxiliary space of the product of $R$ matrices,
$\tau_a^b(u)= \sum_m \prod_{i=1}^LR_{a_i b_i}^{m_i m_{i+1}}(u)$. The $R$ matrices satisfy the Yang-Baxter equations, hence the transfer matrices with different spectral parameters commute, which meant the exact integrability of the system \cite{Zvb}. Then we construct $R$ matrices of different type, related to the initial one by the anticlockwise and clockwise rotations ${\bar R}_{a b}^{m n}(u) = R^{a b}_{n m }(u)$ and ${\tilde R}_{a b}^{m n}(u) = R^{b a}_{m n}(u)$. The transfer matrix ${\bar \tau}(u)$ can be constructed in a way similar to the case of $\tau (u)$. The partition function $Z$ of the considered quantum one-dimensional system 
\begin{equation}
Z = \lim_{N \to \infty} {\rm Tr} \prod_{i=1}^{N/2} \tau(u_i){\bar \tau}(0) \, 
\end{equation}
where $N$ is the Trotter number. It is equal to the partition function of the inhomogeneous classical vertex model with alternating rows on the square lattice of size $L \times N$, $L$ is the size of the chain, with spectral parameters $u_i$ (of order of $N^{-1}$). Consider four-spin interactions on the (classical) two-dimensional lattice with the coupling parameters, which depend on $(N/\beta)^{-1}$, where $\beta$ is the inverse temperature.  
Corresponding column-to-column transfer matrices are known as quantum transfer matrices, which describe transfer in the horizontal direction. The magnetic field $h$ is included via twisted boundary conditions. 
\begin{equation}
\tau_{QTM}(u) = \sum_{\{m_i\}} e^{m_1h/t}\prod_{i=1}^{N/2} R_{a_{2i-1}b_{2i-1}}^{m_{2i-1}m_{2i}}(u-u_i)
{\tilde R}_{a_{2i}b_{2i}}^{m_{2i}m_{2i+1}} \ . 
\end{equation}
We are interested in the properties of the system in the thermodynamic limit $N,L \to \infty$. The quantum transfer matrix has a gap between the largest eigenvalue and the next-largest ones. This is why the free energy of the quantum one-dymensional spin system per site $f$ can be calculated from the largest eigenvalue of the quantum transfer matrix $\Lambda_0(u)$ as $f=-T \lim_{N \to \infty} ln [\Lambda_0(u=0)]$. 

All eigenvalues of the Hamiltonian Eq~(\ref{A1}) can be parametrized by the quantum numbers, rapidities $x_j$, related to the quasimomenta of eigenstates. Rapidities satisfy the Bethe ansatz equations, which can be written as 
\begin{equation} 
\frac{\phi_-(x_j)\phi_+(x_j+2i)}{\phi_+(x_j)\phi_-(x_j-2i)}= e^{2h/t}\frac{Q(x_j+2i)}{Q(x_j-2i)} \ , j=1, \dots M^* \ . 
\end{equation}
with $\phi_+(x) = \prod_{l=1}^{N/2} \sinh [\theta (x+ix_l)/2]$, $\phi_- (x) = \sinh^{N/2} [\theta x/2]$ and $Q(x) = \prod_{j=1}^{M^*} \sinh [\theta (x-x_j)/2]$. The largest eigenvalue of the quantum transfer matrix is related to $M^*=L/2$. The eigenvalue of the quantum transfer matrix is related to the eigenvalue of the row-to-row transfer matrix as $\Lambda_Q (ix) =  \Lambda(x)/\sinh^N (i\theta)$. Let us write $\Lambda(x) = \lambda_1(x) + \lambda_2(x)$ with 
\begin{eqnarray} 
&&\lambda_1(x) = \phi_+(x)\phi_-(x-2i)e^{h/t}\frac{Q(x+2i)}{Q(x)} \ , \nonumber \\
&&\lambda_2(x) = \phi_-(x)\phi_+(x+2i)e^{-h/t}\frac{Q(x-2i)}{Q(x)} \ .
\end{eqnarray}
Then we introduce the auxiliary functions $b (x) = \lambda_1(x+i)/\lambda_2(x+i)$, and ${\bar b} (x) = \lambda_2(x-i)/\lambda_1(x-i)$. One can check that 
\begin{equation}
\Lambda(x+i) =[1+ b(x)]\lambda_2(x+i) \ , \ \Lambda(x-i) =[1+ {\bar b}(x)]\lambda_1(x-i) \ . 
\end{equation}
Then it follows that
\begin{eqnarray}
&&b(x) = e^{2h/t}\prod_{\pm} \frac{\phi_{\pm}(x\pm i)}{\phi_{\pm}(x+2i\pm i)}\frac{Q(x+3i)}{Q(x-i)} \ , \nonumber \\
&&{\bar b}(x) = e^{-2h/t}\prod_{\pm} \frac{\phi_{\pm}(x\pm i)}{\phi_{\pm}(x-2i\pm i)}\frac{Q(x-3i)}{Q(x+i)} \ . 
\end{eqnarray}
These auxiliary functions are analytic, and nonzero. Also the functions $b(x)$ and $1+b(x)$ have constant asymptotic behavior for the strip $-1 < {\rm Im} x \le 0$. The functions ${\bar b}(x)$ and $1 + {\bar b}(x)$  have constant asymptotic behavior for the strip $0 \le {\rm Im} x < 1$. Finally we can denote 
$a(x)= b[(2/\pi)(x+i\epsilon)]$ and ${\bar a}(x) ={\bar b}¯[(2/\pi)(x-i\epsilon )]$ 
with an infinitesimal $\epsilon >0$. Taking the logarithmic derivative of these functions, Fourier transforming the equations, eliminating the functions $Q(x)$, and then inverse-Fourier transforming, and taking the limit $N \to \infty$, one gets two nonlinear integral equations
\begin{eqnarray}
&&\ln a = -\frac{2\pi  \sin (\theta) c(x)}{t} +\frac {h}{t}  + \nonumber \\ 
&&\int_{-\pi}^{\pi} dy g(x-y) \ln (1+a) - \nonumber \\ 
&&\int_{-\pi}^{\pi}dy g(x -y -i[2\theta -\epsilon])\ln (1+{\bar a}) \ , \nonumber \\  
&&\ln {\bar a} = - \frac{2\pi \sin (\theta) c(x)}{t} -\frac{h}{t}  + \nonumber \\
&&\int_{-\pi}^{\pi} dy g(x-y) \ln (1+{\bar a}) - \nonumber \\ 
&&\int_{-\pi}^{\pi}dy g(x -y +i[2\theta -\epsilon])\ln (1+a) \ , \nonumber \\ 
\label{e1}
\end{eqnarray}
with
\begin{equation}
c(x) = \frac{1}{2\theta \cosh(\pi x/\theta)} , 
\end{equation}
and
\begin{equation}
g(x) = \frac{1}{4\pi} \int_{-\infty}^{\infty} d y \frac{\sinh ((\pi -2\theta)y/2)\cos(xy)}{ \cosh (\theta y/2)\sinh ((\pi -\theta)y/2)} \ . 
\end{equation}

For the function $f(x) = -t \lim_{N \to \infty} \ln \Lambda_Q(x)$ one gets 
\begin{equation}
f(ix) = e_0(x) -\frac{t}{2\pi} \int dy \frac{\ln ([1+a(y)][1+{\bar a}(y)])}{\cosh(x-y)} \ . 
\end{equation}
with $e_0(0)$ being the ground state energy of the quantum spin chain per site divided by $J$. The free energy of the quantum spin chain per site divided by $J$ is then
\begin{equation}
f = e_0 -t \int_{-\infty}^{\infty} dx c(x) \ln [(1+a)(1+{\bar a})] \ . 
\label{f1}
\end{equation} 

In the case $|\Delta | > 1$ it is possible to perform similar procedure with $\Delta = \cosh \Phi$ for the antiferromagnetic chain. We get 
\begin{eqnarray}
&&\ln a = -\frac{2\pi  \sinh (\Phi) c(x)}{t} +\frac {h}{t}  + \nonumber \\ 
&&\int_{-\pi}^{\pi} dy g(x-y) \ln (1+a) - \nonumber \\ 
&&\int_{-\pi}^{\pi}dy g(x -y -i[2\Phi -\epsilon])\ln (1+{\bar a}) \ , \nonumber \\  
&&\ln {\bar a} = - \frac{2\pi  \sinh (\Phi) c(x)}{t} -\frac{h}{t}  + \nonumber \\
&&\int_{-\pi}^{\pi} dy g(x-y) \ln (1+{\bar a}) - \nonumber \\ 
&&\int_{-\pi}^{\pi}dy g(x -y +i[2\Phi -\epsilon])\ln (1+a) \ , \nonumber \\ 
\label{e2}
\end{eqnarray}
where   
\begin{equation}
c(x) = \frac{1}{2\pi} \left[\frac{1}{2} +\sum_{n=1}^{\infty} \frac {\cosh (inx)}{\cosh (n\Phi)}\right] \ , 
\end{equation}
and
\begin{equation}
g(x) = \frac{1}{2\pi} \left[\frac{1}{2} + \sum_{n=1}^{\infty} \frac{e^{-n\Phi}\cosh (inx)}{\cosh(n\Phi)} \right] \ . 
\end{equation}
with $\cosh (\Phi) =\Delta$. The free energy per site is 
\begin{equation}
f = e_0 - t \int_{-\pi}^{\pi} dx c(x) \ln [(1+a)(1+{\bar a})] \ .
\label{f2}
\end{equation} 
For the ferromagnetic chain we use the above mentioned connection for the free energy. 

Hence, the solution of the nonlinear integral equations with respect for the functions $a(x)$ and ${\bar a}(x)$ determines the free energy of the spin-1/2 chain with the uniaxial magnetic anisotropy. Then, differentiating the function $f$ with respect to temperature, external magnetic and electric field, and distortion, we obtain thermodynamic characteristics of the spin chain material, necessary for the description of the magneto-caloric, electro-caloric, and elasto-caloric effects in that system. 

To calculate the entropy (the derivative of the Helmholtz free energy with respect to the temperature) it is useful to perform the following trick \cite{Kl}. Namely, let us define the new functions $A=\partial \ln a/\partial t$ and ${\bar A} = \partial \ln {\bar a}/\partial t$. The equations for the functions $A$ and ${\bar A}$ can be obtained from Eqs.~(\ref{e1}) and (\ref{e2}) by analytic differentiation. In the right hand sides of those obtained equations there will be $A/(1+a)$ and ${\bar A}/(1+{\bar a})$. We see \cite{Kl} that the equations for $A$ and ${\bar A}$ are {\it linear} integral equations if one regards the functions of $a$ and ${\bar a}$ as given. Once the integral equations (\ref{e1}) and (\ref{e2}) are solved for $a$ and ${\bar a}$, the integral equations for $A$ and ${\bar A}$ associated with Eqs.~(\ref{e1}) and (\ref{e2}) can be solved. The entropy is determined as the function of $a$, ${\bar a}$, $A$ and ${\bar A}$. The trick permits to avoid numerical differentiation.
For calculation of the specific heat it is possible to introduce functions related to the second derivatives of $\ln a$ and $\ln {\bar a}$ with respect to $T$. Those functions also satisfy linear integral equations.

\section{Limiting cases}
\label{AppendixB}

At high temperatures $T \gg J, J^z$ at $H=0$ one gets for the free energy per site $f = -T\ln 2$, as it must be, so that the entropy in this limit is constant. This is why the specific heat and any expansion coefficients are zero. 

Let us study the case of low temperatures in the absence of the magnetic field. For example, consider the ferromagnetic spin 1/2 chain with the easy-axis magnetic anisotropy $\Delta > 1$. The low temperature free energy of the spin-1/2 chain per site is  \cite{JM}
\begin{equation}
f= -\frac{t^{3/2}}{\sqrt{2\pi}}e^{-(1-\Delta)/t} + \dots \ . 
\label{FreeEnergy1}
\end{equation}
For the easy-axis antiferromagnetic chain one has \cite{JM}
\begin{equation}
f=e_0 -\exp(-B/t)[\sqrt{A}t^{3/2} - \frac{(k^2+k+1)}{4\pi(1-k)^{2}}A^{3/2}t^{5/2} + \dots] \ , 
\label{gapped}
\end{equation}
where 
\begin{eqnarray}
&&A = \frac{k'}{2J K k^2 \sinh \Phi} \ , \nonumber \\
&&B = \frac{Kk'}{\pi}\sinh(\Phi) \ ,
\end{eqnarray}
where the elliptic modules $k$ and $k'$, and the elliptic half-period $K$ are determined via the value 
$q= \exp(-\Phi)$ as 
\begin{eqnarray}
&&K=\frac {\pi}{2} \prod_{n=1}^{\infty}\left[\frac{1+q^{2n-1}}{1-q^{2n-1}}\frac {1-q^{2n}}{1+q^{2n}}\right]^2 \ , \nonumber \\
&&k= 4\sqrt{q}\prod_{n=1}^{\infty}\left[\frac {1+q^{2n}}{1+q^{2n-1}}\right]^4 \ , \nonumber \\
&&k'= \prod_{n=1}^{\infty}\left[\frac {1-q^{2n-1}}{1+q^{2n-1}}\right]^4 \ .
\end{eqnarray}
Hence, the low temperature specific heat and the low temperature expansion coefficients are exponentially small in these regions of $\Delta$.

For the easy-plane spin chain the low temperature part of the free energy can be written as \cite{Zvb} 
\begin{equation} 
f = e_0 - \frac {\pi t^2}{6v} \ , 
\label{gapless}
\end{equation}
with the ground state energy $e_0$ of the easy-plane spin chain and with the velocity of the low energy gapless excitation 
\begin{equation}
v=\frac{\pi \sin (\theta)}{\theta} \  
\end{equation}
for the antiferromagnetic chain and 
\begin{equation}
v=\frac{\pi J\sin (\pi-\theta)}{(\pi -\theta)} \  
\end{equation}
for the ferromagnetic chain. This is why the low temperature entropy of the spin chain per site is ${\cal S} = T/3v$. The low temperature specific heat and the electric expansion coefficient are linear in $T$. At the point $\Delta =1$ the low temperature entropy of the antiferromagnetic spin-1/2 chain as a function of the temperature has a jump. Hence, the low temperature specific heat manifest the feature at that point. Low temperature expansion coefficients reveal the behavior, similar to the one of the specific heat. 

For the isotropic antiferromagnetic chain one takes the limit $\theta \to 0$ ($v = \pi/2$) and obtains (taking into account small logarithmic correction \cite{Kl}
\begin{equation}
f=e_0 -\frac{\pi t^2}{6v} \left(1+ \frac{3}{8\ln^3(\pi/t)}\right) \ . 
\end{equation}

\end{document}